\documentstyle[twoside,fleqn,espcrc2]{article}

\newcommand{\AmS}{{\protect\the\textfont2
  A\kern-.1667em\lower.5ex\hbox{M}\kern-.125emS}}

\title{Solitons, Black Holes and Duality in String Theory}

\author{Ramzi R. Khuri\address{Physics Department, 
McGill University, Montreal, Quebec, Canada H3A 2T8}
\thanks{Based on lectures given at Asia-Pacific Center 
for Theoretical Physics, Seoul, Korea, 5-10 November 1996, 
33rd Karpacz Winter School of Theoretical Physics, Karpacz, 
Poland, 13-22 February 1997 and Humboldt University of
Berlin, 21-24 February 1997. To appear in Nuclear Physics
B - Proc. Suppl.
Research supported by NSERC of Canada and Fonds FCAR du 
Qu\' ebec. McGill/97-4, hep-th/9704110.}}
              
\begin{document}

\begin{abstract}
These lectures are intended as an introduction to
some of the basic aspects of string solitons,
duality and black holes. We begin with a discussion of the 
role of
classical solutions in duality, then focus on string/string 
duality 
and fundamental 
membranes. Finally, we examine the feature of compositeness of 
string
solitons, and its implications for bound states and black hole 
thermodynamics. As these lectures are aimed primarily at those
less familiar with this field, technical details are minimized.  
\end{abstract}

\maketitle

\section{INTRODUCTION}

The standard model of elementary particle physics has proven 
very succesful 
in describing three of the four fundamental forces of nature. 
In the most
optimistic scenario, the standard model can be generalized to 
take the
form of a grand unified theory (GUT), in which quantum 
chromodynamics, 
(QCD) describing the strong force, and the electroweak theory, 
unifying
the weak interaction with electromagnetism, are synthesized 
into a single
theory in which all three forces have a common origin. 

The framework for
studying these three forces is that of Yang-Mills gauge theory,
a certain class of quantum field theory based on the principle 
of gauge
symmetry. In any quantum-mechanical theory, the natural length 
scale
associated with a particle of mass $m$ (such as an elementary 
particle)
is given by the Compton wavelength $\lambda_C=\hbar/mc$, where 
$\hbar$
is Planck's constant divided by $2\pi$ and $c$ is the speed of 
light.
Scales less than $\lambda_C$ are therefore
unobservable within the 
context of 
the quantum mechanics of this elementary particle.

Quantum field theory, however, has so far proven unsuccessful in 
describing
the fourth fundamental force, namely gravitation. The succesful 
framework  
in this case is that of general relativity, which, however, 
does 
not
seem to lend itself to a straightforward attempt at 
quantization. 
The 
main problem in such an endeavour is that the divergences 
associated with
trying to quantize gravity cannot be circumvented (or 
``renormalized")
as they are for the strong, weak and electromagnetic forces. 

Among the most interesting objects predicted by general 
relativity are 
black holes, which represent the endpoint of gravitational 
collapes. 
According to relativity, an object of mass $m$ under the 
influence of only 
the 
gravitational force ({\em e.g.} neutral with respect to the 
other three 
forces) will collapse into a region of spacetime bounded by a 
surface,
the event horizon, beyond which signals cannot be transmitted 
to an outside 
observer. The event horizon for the simplest case of a 
static, spherically
symmetric black hole is located at a radius $R_S=2 G_N m/c^2$, 
the 
Schwarzschild radius, from the collapsed matter at the 
center of the sphere,
where $G_N$ is Newton's constant. 
      
\def\buildchar#1#2#3{{\null\!
   \mathop{\vphantom{#1}\smash#1}\limits%
   ^{#2}_{#3}%
   \!\null}}
% $\buildchar{\sim}{<}{}$
\def\ha{{1\over 2}}
  
In trying to reconcile general relativity and particle 
physics, even at an
intuitive level, a natural question to ask is whether 
they have a common
domain. This would arise when an elementary particle 
may exhibit features 
associated with gravitation, such as an event horizon. 
This may occur 
provided 
$\lambda_C \ { } 
\lower3pt\hbox{$\buildchar{\sim}{<}{}$} \ { } R_S$,
which implies that, even 
within the framework of quantum mechanics, 
an event horizon for an 
elementary
particle may be observable. Such a condition is equivalent to
$m \ { } \lower3pt\hbox{$\buildchar{\sim}{>}{}$} \ { } m_P=
\sqrt{\hbar c/G} \sim 10^{19} GeV$, 
the Planck mass, or 
$\lambda_C \ { } 
\lower3pt\hbox{$\buildchar{\sim}{<}{}$} \ { } l_P
=\sqrt{\hbar G_N/c^3}$,
the Planck scale. 
It is in this domain that one may study a theory
that combines quantum mechanics and gravity, the so-called
{\em quantum gravity} (henceforth we use units in which
$\hbar=c=1$).

At the present time, string theory, the theory of one-dimensional
extended objects, is the only known reasonable
candidate theory  of quantum gravity. The divergences inherent 
in trying to quantize point-like gravity seem not to arise in
string theory. Furthermore, string theory has the potential
to unify all four fundamental forces within a common framework.

The two-dimensional worldsheet swept out by a string is embedded
in a higher-dimensional (target) spacetime, which in turn 
represents a 
background
for string propagation. At an intuitive level, one can see 
how point-like
divergences may possibly be avoided in string theory by 
considering
the four-point amplitudes arising in string theory 
(see, {\em e.g.},
\cite{GSW}). Unlike those of field theory,
the four-point amplitudes in string theory do not have 
well-defined
vertices at which the interaction can be said to take place,
hence no corresponding divergences associated with the zero size
of a particle. A simpler way of saying this is that the finite 
size of 
the string smooths out the divergence of the point particle.

The ground states of string theory correspond to conformal 
invariance
of the two-dimensional sigma model of a genus zero (sphere) 
worldsheet.
Solving the beta-function equation of this sigma-model then 
correspond
to classical solutions of string theory. Within this classical 
theory,
the perturbative parameter is $\alpha'=1/(2\pi T_2)=l_s^2$, 
where $T_2$ is
the tension of the string and $l_s$ is the string length.

Perturbative quantum corrections in string theory take the 
form of 
an expansion in the genus of the worldsheet, with coupling 
parameter 
$g=\exp{\phi}$, where $\phi$, the dilaton, is a dynamical scalar
field.

Consistent, physically acceptable string theories possess 
supersymmetry
between bosons and fermions, and supersymmetric string theory 
require
a ten-dimensional target space. This leads to another feature 
of string
theory, namely, compactification, {\em i.e.}, the splitting 
of the
ten dimensional vacuum into the product of
a four-dimensional vacuum and
a compact six-dimensional manifold which may be shrunk to a 
point. 

Finally, the property of string theory that is of most interest 
to us in these lectures is that of {\em duality}. At the 
simplest
level, duality is a map that takes one theory into another 
theory (or possibly
the same theory in a different domain). An immediate consequence 
of duality
is that the two theories are physically equivalent in the 
appropriate
domains. It then follows that calculations performed on one side
can be immediately carried over into the other, even if direct
calculations in the latter theory may have previously been 
intractable.

The two most basic dualities in string theory are target 
space duality, $T$-duality,
(see \cite{Tduality} and references therein) 
and strong/weak coupling duality, $S$-duality
(see \cite{Sduality} and references therein). 
Suppose
in a compactification one of the dimensions of the 
six-dimensional
compactification manifold is wrapped around a circle with radius 
$R$. Then $T$-duality is a generalization of a map that
takes a string theory with radius
$R$ into a theory with radius $\alpha'/R=l_s^2/R$. This 
implies that
a radius smaller than the string scale is equivalent to a radius
larger than the string scale. Effectively, then, the string 
scale
is a minimal scale, which conforms to our previous intuition 
that
the size of the string smooths out the point-like divergence. 
$T$-duality is a classical, 
worldsheet duality and in various compactifications has been 
shown
to be an exact duality in string theory. 

$S$-duality, by contrast, is a quantum (string loop), spacetime 
duality and  generalizes the map that takes the string coupling 
constant $g$ to its inverse $1/g$. Such a map takes the weak 
coupling 
domain into the strong coupling domain within a given string 
theory 
and allows us to use perturbative results in the latter. Also 
unlike
$T$-duality, $S$-duality has only been established exactly in 
the 
low-energy limit of string theory. 

These
two dualities and the interplay between them
are at the heart of the recent activity in string theory.
This activity has also been fueled by the realization that 
perturbative string theory is insufficient
to answer the most fundamental questions of string theory, such 
as vacuum selection, supersymmetry breaking, the cosmological
constant problem and, finally, the problem of understanding
quantum gravity from string theory. All these questions require
{\em nonperturbative} information. 

What kind of objects arise in nonperturbative physics? 
Solitons, or topological defects, (see \cite{Raj}\ and 
references
therein), are inherently nonperturbative solutions, representing
objects with mass $m_s \sim 1/g^2$, where $g$ is the coupling
constant of the theory. Examples of solitons are magnetic 
monopoles or domain walls. The connection between duality and 
solitons often involves the interchange of perturbative,
fundamental (electric) particles with nonperturbative, solitonic
(magnetic) objects. This is the main feature of the 
Montonen-Olive
conjecture \cite{Mont}\ for $N=4$, $D=4$ supersymmetric
Yang-Mills gauge theory, which postulates the existence of a 
dual version of the theory in which electric gauge bosons
and magnetic monopoles interchange roles. In this scenario,
the monopoles become the elementary particles and the
gauge bosons become the solitons. 

Since string theory contains various higher-rank tensor fields 
in
higher dimensions, solitons in general take the form of
higher-dimensional extended objects, the so-called 
{\em $p$-branes}
(see \cite{Prep}\ and references therein).
A $p$-brane sweeps out a $p+1$-dimensional worldvolume, 
generalizing
the two-dimensional worldsheet of the string ($1$-brane). In 
what follows, we will discuss some of the main features of 
these
solutions of string theory and their importance in establishing
dualities. At the heart of our discussion will be the connection
between elementary and solitonic solutions of string theory with
perturbative and nonperturbative states of string theory, a 
connection
which has lead to many interesting
results in the study of duality and black
holes in string theory, but which is also still
not completely 
understood. 

In the rest of the first lecture, we summarize the elementary
string and solitonic fivebrane solutions in ten dimensions,
and briefly discuss string/fivebrane duality and generalizations
of $p$-brane dualities. In the second lecture, we focus on the
case of string/string duality and discuss the role played by 
duality
in requiring the inclusion of a certain class of fundamental 
membranes
(D-branes). We also comment on how $p$-branes may shed light 
on
dualities involving eleven-dimensional supergravity. Finally, 
in the
third lecture, we apply our knowledge of $p$-branes and duality
to try to test string theory as a  
theory of quantum gravity. We do this
by first establishing the compositeness 
feature
of string solitons to form a bound state picture of black holes.
We then combine this result with the $p$-brane/string state 
connection
to obtain results in (classical GR) black hole thermodynamics
from a microscopic counting of (quantum) string states.

\section{CLASSICAL SOLUTIONS AND DUALITY}

Let us consider a specific ten-dimensional
superstring theory, the heterotic string
theory, which is the combination of a right moving superstring 
and
a left moving bosonic string \cite{Gross}. A classical 
solution of heterotic string theory corresponds to 
superconformal
invariance of the worldsheet sigma-model. The resulting
beta-function equations are then equivalent to the equations
of motion of an effective action, in which the worldsheet
couplings
$G_{MN}$, $B_{MN}$ and $\phi$ are interpreted 
as background spacetime fields, namely the metric, antisymmetric
tensor and dilaton, respectively. In the low-energy limit,
the ten-dimensional, bosonic, effective action in the gravity
sector is given by

\begin{equation}
I_{10}=\int d^{10}x \sqrt{-g} \left( R - {1\over 2}
(\partial\phi)^2 - {1\over 12}e^{-\phi} H_3^2 \right)
\end{equation}
where $g_{MN}=G_{MN}\exp{(-\phi/2)}$ is the canonical 
(or Einstein)
metric and $H_{MNP}=\partial_{[M}B_{NP]}$. This is simply the 
bosonic
sector of $N=1$, $D=10$ supergravity. 

An interesting question to pose is the following: does there
exist a solution to the equations of motion
of $I_{10}$  representing the exterior of a
(macroscopic) straight, fundamental string? Such a solution
should have a $P_2 \times SO(8)$ symmetry, the $P_2$
representing Poincar\' e invariance for the string and the 
$SO(8)$ the maximal spatial symmetry in the rest of the
space, transversal to the string. 

It turns out that the answer to this question is "No". One
can find a solution valid everywhere except at the origin,
where the string is located. There one has to compensate
a delta-function singularity by introducing a source term,
in analogy with the source charge one must introduce to
account for electric charge in electrostatics:
$\vec\nabla \cdot \vec E=4\pi\rho$. In string theory,
a natural source term for a string requires the addition
of a sigma-model action

\begin{eqnarray}
S_2={T_2\over 2} \int d^2 \sigma \Bigl( \sqrt{-\gamma} \gamma^{ij}
\partial_i X^M \partial_j X^N G_{MN} & \\
+  \epsilon^{ij}
\partial_i X^M \partial_j X^N B_{MN} & \Bigr), \nonumber
\end{eqnarray}
where $i,j$ are worldsheet indices, $M,N$ are spacetime 
indices and $\gamma_{ij}$ is the worldsheet metric. Then the 
combined action $I_{10} + S_2$ does indeed have the
desired string-like solutions. These are given by \cite{Dab}   

\begin{eqnarray}
ds^2=& e^{3\phi/2} (-dt^2+dx_1^2)+e^{-\phi/2} d \vec y \cdot
d \vec y \\
e^{-2\phi}=&1+k_2/y^6, \qquad\qquad B_{01}=
\pm e^{2\phi},\nonumber
\end{eqnarray}
where $x_1$ is the direction of the string, $\vec y$ is a 
vector in 
the eight-dimensional transverse space $(23456789)$ and $y$ is 
the
radial coordinate in this space. 

Notice that this solution is
electrically charged under the field $B_{MN}$, which is the 
string-like analog to the gauge potential $A_M$ of an 
electromagnetic
point particle in four dimensions. For
the latter case, electric charge is given by 
$\int_{S^2} \tilde F_2$, where $S^2$ is a two-sphere 
surrounding the charge and where $\tilde F_2$, the {\em dual}
of the electromagnetic field strength $F_2$, is given by 
$\tilde F_{MN}=\ha \epsilon_{MNPQ} F^{PQ}$, where  
$\epsilon_{MNPQ}$
is the totally antisymmetric Levi-Civita tensor. 
For the string,
the electric charge is given by $k_2 \sim \int_{S^7} e^{-\phi}
\tilde H_3= \int_{S^7} K_7$, where $K_7$ is a 
seven-tensor,
since the spacetime is ten-dimensional and the sphere 
surrounding
the string is a seven-sphere. Also note that as we approach
the singularity at $y=0$, the string coupling 
$g=\exp{\phi} \to 0$,
so that this solution may reasonably correspond to the exterior 
solution of a perturbative string state with charge under the
(axionic) field $B_{MN}$.

Now another question arises: what is the  coresponding magnetic
object, {\em i.e.} the one dual to the string? Again let us take
the analogy with four-dimensional point particles.
There the magnetic charge of a monopole is given by 
$\int_{S^2} F_2$, since in general $\tilde{(\tilde F)} = \pm F$.
So for the string we would expect a nonzero charge proportional
to $\int_{S^3} H_3$. Such an object would then have a field
strength $K_7$, and so would couple to a six-tensor potential
$A_6$, where $K_7=dA_6$. An object with a six-dimensional
worldwolume is called a {\em fivebrane}, since it has five
spatial directions (and one time). So here we look for a 
solution
with $P_6 \times SO(4)$ symmetry, but consider the action 
$I_{10}$
alone, since we expect this object to be a soliton, hence 
nonsingular.
This turns out to be the case, and we get the solution 
\cite{FB} 

\begin{eqnarray}
ds^2=& e^{-\phi/2} (-dt^2+d\vec x \cdot d\vec x)+
e^{3\phi/2} d \vec y 
\cdot
d \vec y \\
e^{2\phi}=&1+k_6/y^2, \qquad\qquad H_3=
\pm 2k_6\epsilon_3\nonumber
\end{eqnarray}
where here $\vec x$ is a vector in the five-dimensional 
space $(12345)$ in
which the fivebrane extends and $\vec y$ is a vector in the 
four-dimensional
transverse space $(6789)$. $\epsilon_3$ is the volume form 
on the unit
three-sphere. This solution, in contrast to the string, is 
nonsingular,
but only when written in the string sigma-model frame 
(with metric
$G_{MN}$). It is also clearly nonperturbative, since as we 
approach
$y=0$, the string coupling $g\to \infty$, and so would 
correspond to
a nonperturbative (magnetic) state of the string theory.

We now have a fundamental object (the string) and a solitonic 
object
(the fivebrane). Suppose we were to follow Montonen and Olive 
and
postulate the existence of a dual theory, in which the roles of
these objects were interchanged. In our original theory, 
heterotic
string theory, the string solution is fundamental (electric), 
perturbative 
and singular, while the fivebrane solution is solitonic 
(magnetic),
nonperturbative and nonsingular. In the dual theory, a 
{\em heterotic
fivebrane} theory, the fivebrane is fundamental, perturbative 
and
singular, while the string is solitonic, nonperturbative and 
nonsingular.

But how can this be, since we said that the string is singular 
and
the fivebrane nonsingular in the string sigma-model frame? 
In the 
low-energy
limit, we can rewrite the effective action $I_{10}$ in terms 
of $K_7$.
However, in order to regard the fivebrane as singular, we need 
to couple
a fivebrane source. Proceeding in analogy with the string and 
taking
into account scaling considerations, one obtains a fivebrane 
sigma-model
action $S_6$
with six-dimensional worldvolume, with worldvolume metric 
$\tilde\gamma_{ij}$  
and spacetime metric $\tilde G_{MN}=g_{MN} \exp{(-\phi/6)}$. 
In terms 
of this metric, the fivebrane solution is indeed singular and 
perturbative
with respect to the fivebrane coupling 
$\tilde g=\exp{(-\phi/3)}$.
 Here
one adds $S_6$ as a source to balance the delta-function 
singularities,
and the charge of the fivebrane is electric with respect to 
the
field $A_6$.   
In this frame, the string arises as a nonsingular, magnetic
and nonperturbative solution.

Of course this picture is only complete in the low-energy 
limit, since 
we don't really know what the fivebrane theory is (recent 
progress
on this issue will be discussed in other lectures
later in this volume). 
In particular,
direct quantization of the fivebrane is far from clear, and 
many of 
the nice features of strings (such as conformal invariance) are
clearly absent in any naive attempt to understand fivebranes.
Note also that these solutions are not solitons in the 
conventional
sense of having a (manifestly) finite extent. A background
in this case is regarded as nonsingular if the specific probe
of the theory cannot reach it in finite proper time \cite{Sing}. 
For example, the fivebrane is a nonsingular solution of 
string theory 
because a test string probe never sees a singularity at the 
origin,
where the fivebrane is located.
 
As solutions, both the string and fivebrane preserve half the 
supersymmetries
of the low-energy supergravity theory, and both can be 
generalized to
multi-soliton solutions. In analogy with magnetic monopoles, 
this 
extension follows from the ``zero-force'' condition that arises 
as
the result of the cancellation of the attractive 
gravitational and dilatational forces against the repulsive 
antisymmetric
tensor force (generalizing the electrostatic force). This 
condition
is intimately tied to the the existence of supersymmetry, 
which results
in the saturation of the so-called Bogomol'nyi bound
\cite{Bog}.
In fact, both solutions may be regarded as extremal limits of 
two-parameter
black string or black fivebrane solutions \cite{Hors}, 
in which the 
extremality
bound is saturated so that both solutions are stable. 

Duality between $p$-branes can be generalized to arbitrary 
higher
dimensional objects in arbitrary spacetime dimemsions
\cite{DLgen}, if we 
put
aside issues of existence and quantization of higher-membrane 
theories.
A $p$-brane with worldvolume dimension $d=p+1$ has a rank $d+1$
field strength. The dual field strength in $D$-dimensions then 
has 
rank $D-d-1$. This implies that the dual object is a 
$\tilde p$-brane
with worldvolume dimension $\tilde d=\tilde p +1$ such that
$\tilde d +1 = D-d-1$. It then follows that $d+\tilde d +2=D$, 
or
$p+\tilde p +4=D$ (for string/fivebrane duality in $D=10$, $p=1$
and $\tilde p=5$). The same consequences of string/fivebrane
duality, namely, fundamental/solitonic (or electric/magnetic),
perturbative/nonperturbative and singular/nonsingular interchange
then follow, with Dirac-type duality relations between charges, 
couplings
and $p$-brane tensions. Both elementary and solitonic $p$-branes
also in general arise as extremal limits of general black 
$p$-brane
solutions. A feature of $p$-brane solutions that will be 
important to us
in the third lecture is that of compositeness, {\em i.e.} 
the ability
to construct general $p$-brane solutions as composites of the 
basic
fundamental and solitonic building blocks.

\section{STRING/STRING DUALITY AND FUNDAMENTAL MEMBRANES}

In this lecture we focus on the case of string/string duality.  
One obvious reason why we would want to do so is that, since
string theory is much better understood than other membrane
theories, we can compare two theories we already understand
that are related via duality (or a single theory in different 
domains).
Another reason to single out string/string duality is that, 
in some
sense, we would like a theory dual to string theory to have 
the basic 
features of string theory, such as conformal invariance.

String/string duality was first considered in the context of 
six-dimensional heterotic/heterotic duality toroidally reduced 
to four
dimensions \cite{stst}. An interesting and nontrivial consequence
of this reduction is that, in the (effective) four-dimensional
string/string duality, the spacetime,
strong/weak coupling $S$-duality of one version goes into
the 
worldsheet, target space $T$-duality of the dual version. 
Since $T$-duality is established 
for this type of compactification, the conjectured $S$-duality
would then follow as a consequence of string/string duality.  

More recently, interest has focused on duality between different
string theories. In particular, the conjecture of
duality was made between heterotic string theory compactified
on $T_4$ and type IIA string theory compactified on $K3$,
a four-dimensional manifold which in the orbifold limit looks 
like
$T_4/Z_2$ \cite{hult,ed}. In this case, one has in some sense a 
more
straightforward strong/weak coupling duality relating the two
theories with $g_{IIA}=1/g_{het}$. Evidence for this duality 
can be
seen from considerations of low-energy
field-content, supersymmetry,
spectra of states and even one-loop tests. Finally,
from the point of view of $p$-brane solutions, the usual 
interchange
of fundamental and solitonic solutions (both strings in this
case) occurs. In particular, the fundamental heterotic string
appears as a soliton of the type $IIA$ theory and vice-versa
\cite{shs}.

Consider now an arbitary $p$-brane solution of either of these
dual theories. Let us study the behaviour of this $p$-brane 
under
the duality map. If the $p$-brane is singular in both theories, 
we may discard it as unphysical. If it is nonsingular 
(solitonic)
in both theories, then it can be included in both solitonic 
spectra.
However, if the $p$-brane is singular in one version and 
nonsingular 
in the other, then it can neither be ignored as unphysical nor 
simply
included in a soliton spectrum. In that case, the $p$-brane 
must be 
regarded as a fundamental source. This is the case, of course,
for each of the fundamental strings (heterotic and type IIA). 
As it turns out, however, there also exist membrane solutions
which are nonsingular in heterotic string theory but singular
in type IIA \cite{jkkm}. What is the interpretation of such 
solutions? Another puzzle lies in the peculiar $1/g$ dependence
of the mass of these membranes, intermediate between the
coupling-independent mass behaviour of fundamental $p$-branes
and the $1/g^2$ mass behaviour of solitonic $p$-branes, the 
latter
being typical solitonic behaviour in field theory.

The existence of these membranes, coupled with the assumption
of duality, has pointed to a gap in the formulation of 
$IIA$ string theory, since it appears we must add fundamental 
membranes to the spectrum. This puzzle was resolved in 
\cite{joe},
where it was shown that Dirichlet-branes, or D-branes, 
extended objects defined by mixed Dirichlet-Neumann boundary 
conditions  for {\em open} strings, should
be coupled to types IIA, IIB and type I (open string)
theories and identified
as states carrying both electric and magnetic Ramond-Ramond (RR)
charge. As D-branes will be 
extensively covered later in this 
volume (see also \cite{pcj}\ and references therein for a review 
on D-branes), 
we will not discuss them in detail here. Suffice to say
that this discovery solved two problems at once: the inclusion
of RR states in type II string theory, previously
unaccounted for in the perturbative spectrum, and the 
interpretation
of RR charged solitons arising in the context of string duality.
The mass of a D-brane state (or a D$p$-brane solution) scales 
as $1/g$

 A picture then emerges in which fundamental and solitonic
$p$-branes carrying Neveu-Schwarz-Neveu-Schwarz (NSNS) charge
correspond to perturbative and nonperturbative BPS states of
string theory, while $p$-branes carrying RR charge correspond
to D-brane BPS states which must be coupled to the perturbative
spectrum.

Since type IIA supergravity is the low-energy limit of IIA 
string
theory, and also arises as the Kaluza-Klein reduction of 
eleven-dimensional
supergravity, a natural question to ask is whether the latter 
is
the low-energy limit of a fundamental eleven-dimensional 
theory, the
so-called {\em M-theory} \cite{ed,MTh}. 
The eventual construction of M-theory should lead
to the establishment of the various string/string dualities.
In this framework, the five seemingly distinct string theories
arise as weak coupling limits of the various compactifications 
of the
eleven-dimensional M-theory. 

While we still do not know exactly
what M-theory is (recent progress in this endeavour will be
discussed in other lectures in this volume), we know that
M-theory contains membranes and fivebranes. From the point of
view of $p$-brane solutions, these are represented by a 
fundamental
membrane \cite{Memb}\ and a solitonic fivebrane \cite{Guven}. 
In compactifying to lower dimensions, the membrane and fivebrane
are wrapped or simply reduced to yield the various $p$-branes 
carrying
both NSNS and RR charge. For example, in the Kaluza-Klein (KK) 
reduction
from eleven to ten dimensions (IIA supergravity), 
we obtain the NSNS fundamental string 
(and then dualize to obtain the fivebrane) by  wrapping the 
membrane
around the KK direction, while we obtain the RR membrane in ten 
dimensions
by simply reducing the membrane and the RR point particle from 
the KK
reduction of the metric.

Let us now look at an example of how $p$-branes can shed light
on a duality involving M-theory. Following \cite{Ferrara},
let us consider the conjecture that M-theory compactified on
a particular Calabi-Yau threefold is dual to heterotic string
theory compactified on $K3\times S^1$.     

Point-like (electric) states are obtained in $D=5$ by wrapping 
the
membrane from M-theory around two-cycles in
the Calabi-Yau space. Denote two-cycles and four-cycles 
respectively
by $C^{2\Lambda}$ and $C_{4\Lambda}$, where
$\Lambda=1,...,h_{(1,1)}$.
The charges of these states are obtained from the charge of the
membrane by

\begin{equation}
e_\Lambda=\int_{C_{4\Lambda}\times S^3} G_7,
\end{equation}
where $G_7=\delta {\cal L}/\delta F_4$, where $F_4=dA_3$ is the
field strength of the three-form antisymmetric tensor field.

String-like (magnetic) states in $D=5$ arise by wrapping the
fivebrane
around four-cycles in the Calabi-Yau space. The charges of these
states
are then obtained from the charge of the fivebrane by

\begin{equation}
m_\Lambda=\int_{C^{2\Lambda}\times S^2} F_4.
\end{equation}

Since the membrane and fivebrane are electric/magnetic duals in
eleven
dimensions, the above point-like and string-like states are 
dual to
each
other in the electric/magnetic sense and correspond to 
point-like and
string-like soliton solutions. 
For the specific Calabi-Yau manifold
$X_{24}(1,1,2,8,12)$ with $h_{(1,1)}=3$ and $h_{(2,1)}=243$, 
we match
these point and string solutions/states with
perturbative and nonperturbative solutions/states of
heterotic string theory compactified on $K3\times S^1$, 
in which the gauge symmetry is completely Higgsed.

{}From the ten-dimensional point of view,
the heterotic string compactified on $K3\times S^1$ has the
perturbative
fundamental string state with charge
\begin{equation}
\tilde m_0=\int_{K3\times S^1\times S^2} H_7,
\end{equation}
where $H_7=e^{-\phi} \tilde H_3$, $H_3$ is the field strength 
of the
two-form
antisymmetric tensor field and $\phi$ is the ten-dimensional 
dilaton.
Here the string is not
wrapped around the $S^1$
and so remains a string in $D=5$. The corresponding classical
solution is simply the fundamental string in $D=5$.
 This state is associated with the $B_{\mu\nu}$ field 
and is
dual to a vector in $D=5$.
The string theory also possesses
a perturbative electrically charged
point-like $H$-monopole state
(dual to the magnetically charged $H$-monopole state
of \cite{Hmono}) with charge
\begin{equation}
\tilde e_1=\int_{K3\times S^3} H_7.
\end{equation}
In this case, the string is wrapped around the $S^1$ and
becomes a point in $D=5$.
This state is associated
with the $B_{\mu 5}$ field, where $x_5$ is the direction 
compactified on the $S^1$.   
 The $T$-dual electrically charged
point-like
Kaluza-Klein state has charge $\tilde e_2$ associated with the
$g_{\mu 5}$ field. In this case, the
corresponding
electrically charged solution is given by the extremal 
Kaluza-Klein
black hole solution of heterotic string theory \cite{KK}.
The fundamental
string state can be identified with one of the three states 
arising
from
the M-theory fivebrane by setting $m_0=\tilde m_0$, 
while the $H$-monopole and Kaluza-Klein
states
can be identified with two of the three states arising from the
M-theory
membrane by setting $e_1=\tilde e_1$ and $e_2=\tilde e_2$.
The masses of the identified M-theory 
and heterotic solutions/states also agree,
whether computed from central charge/supergravity
considerations or from the ADM mass of the classical
solutions \cite{Ferrara}.

The dual case is similar:
the heterotic fivebrane wrapped around $K3\times S^1$ has the
nonperturbative
(from the string point of view) point-like state with charge

\begin{equation}
\tilde e_0=\int_{S^3} H_3
\end{equation}
 Here the classical
solution is
simply the heterotic fivebrane wrapped around $K3\times
S^1$,
and which is dual to the fundamental heterotic string.
One also gets from the heterotic
fivebrane a nonperturbative magnetically charged
string-like $H$-monopole state with charge

\begin{equation}
\tilde m_1=\int_{S^1\times S^2} H_3,
\end{equation}
where in this case the fivebrane is
wrapped around the $K3$ but reduced on the $S^1$.
The solution in this case is the usual magnetically
charged $H$-monopole (under the field
$B_{\mu 5}$) \cite{Hmono}, which in $D=5$ is a string.
The $T$-dual magnetically charged string-like
Kaluza-Klein state has charge $\tilde m_2$ under the field
$g_{\mu 5}$. 
The point-like state can be identified with one of the three 
states
 arising from the M-theory membrane by setting
$e_0=\tilde e_0$, 
while the
string-like $H$-monopole and Kaluza-Klein states
can be identified with two of 
the three states arising from the M-theory fivebrane by
setting $m_1=\tilde m_1$ and $m_2=\tilde m_2$. Again,
the M-theory and heterotic masses agree as well.

Note that each of the
three pairs of electric/magnetic dual states obey Dirac 
quantization
conditions. Note also that neither the membrane nor the 
fivebrane
from
M-theory is in itself sufficient to reproduce the perturbative
spectrum
of either the five-dimensional string or the dual
 five-dimensional
point-like
object. This becomes clear when one realizes that, from the
M-theory side,
the membrane wrapped around a two-cycle yields only point-like
states, while
the fivebrane wrapped around a four-cycle yields only 
string-like
states. On
the other hand, from the heterotic compactification,
both the string and point-like theories in $D=5$
contain both string and point-like objects in their 
perturbative
spectra.
In particular, it follows that the $D=5$ spectrum of Calabi-Yau
string
solitons yields the fundamental string states on the heterotic 
side
as well
as the nonperturbative heterotic string states obtained by 
wrapping
the
heterotic fivebrane on $K3$.

One-loop calculations provide further evidence for this duality
\cite{Ferrara}. From anomaly considerations, a
five-dimensional string action arises which is chiral
on the worldsheet. This is especially interesting, since it 
implies
that M-theory
calculations may be carried out in the more familiar setting of
string theory. Further reduction to $D=4$
yields dual $N=2$ supersymmetric theories \cite{SW}.

\section{BOUND STATES AND BLACK HOLE THERMODYNAMICS}

The basic fundamental and solitonic $p$-branes preserve
half the spacetime supersymmetries, and arise
as extremal limits of more general, non-supersymmetric black
$p$-brane
solutions of string theory. It turns out, however, that the
low-energy
supergravity equations of motion possess a
feature that allows for the immediate construction of composite
solutions from the basic ones. Consider, for example, the
fivebrane solution of eq. (4). Then the generalized connection
$\Omega_\mu^{ab}=\omega_\mu^{ab} -(1/2) H_\mu^{ab}$, 
obtained by adding the three-form field strength as torsion 
to the standard metric connection in the sigma-model frame,
is an instanton \cite{Hmono,Inst}: $\Omega_\mu^{ab}=\mp (1/2) 
\epsilon^{ab}{}_{cd} \Omega_\mu^{cd}$.
 The (anti) self-duality of $\Omega$ imposes
a chirality choice on the supersymmetry-generating spinor
$\epsilon$ which results in halving the
number of supersymmetries preserved.  

\def\sqr#1#2{{\vbox{\hrule height.#2pt\hbox{\vrule width
.#2pt height#1pt \kern#1pt\vrule width.#2pt}\hrule height.#2pt}}}
\def\Box{\mathchoice\sqr64\sqr64\sqr{4.2}3\sqr33}

It it not too difficult to see that, by inserting another,
independent instanton, we can construct another solution
with half again of the number of supersymmetries preserved
\cite{bifb}
\begin{eqnarray}
\phi=&\phi_1 + \phi_2,\quad g_{\mu\nu}=\eta_{\mu\nu}\quad   
\mu,\nu=0,1,\\
g_{mn}=&e^{2\phi_1}\delta_{mn}\qquad m,n=2,3,4,5,\nonumber \\
g_{ij}=&e^{2\phi_2}\delta_{ij}\qquad i,j=6,7,8,9,\nonumber \\
H_{mnp}=&\pm 2\epsilon_{mnpq}\partial^q\phi
\qquad m,n,p,q=2,3,4,5,\nonumber \\
H_{ijk}=&\pm 2\epsilon_{ijkl}\partial^k\phi
\qquad i,j,k,l=6,7,8,9, \nonumber 
\end{eqnarray}
with $e^{-2\phi_1}\Box_1 e^{2\phi_1}=e^{-2\phi_2}\Box_1
e^{2\phi_2}=0$,
where $\Box_1$ and $\Box_2$ are the Laplacians in the $(2345)$ 
and
$(6789)$
spaces, respectively.
For constant chiral spinors
$\epsilon=\epsilon_2 \otimes \eta_4 \otimes \eta_4'$, this
double-instanton string solution solves
the supersymmetry equations with zero background fermi fields. 
Due
to the presence of two independent instantons in $\Omega$,
the chiralities of the spinors $\epsilon_2$, $\eta_4$ and 
$\eta_4'$
are correlated by
$(1 \mp \gamma_3)\epsilon_2=(1 \mp \gamma_5)\eta_4
=(1 \mp \gamma_5)\eta_4'=0$,
so that $1/4$ of the spacetime supersymmetries is preserved.

For either $\phi_1=0$ or $\phi_2=0$ we recover the solitonic
fivebrane, which preserves $1/2$ the spacetime supersymmetries,
since
only one instanton is present. In this respect, this string
is the composite of two independent fivebranes intersecting
along the string. This feature can in fact be generalized to
arbitrary
$p$-brane solutions \cite{iarkady,iklaus,ijerome,imirjam},
whereby
a given $p$-brane soliton can be interpreted as the 
intersection of
one or more maximally supersymmetric basic fundamental or 
solitonic
$p$-branes. From the viewpoint of $M$-theory, the statement 
then
translates into saying that all $p$-brane solutions arise as
intersections of membranes and fivebranes \cite{iarkady}.

Compositeness was previously seen in the specific
context of string-like solutions of toroidally compactified 
$D=4$
heterotic string theory \cite{dfkr}, where each solution could 
be
understood
in terms of four independent harmonic functions. The existence 
of
these
latter solutions also pointed to an interesting interplay 
between
supersymmetry and duality. In particular, the different 
supersymmetry
breaking patterns of the string-like solutions conform to the
different large duality groups (containing both $S$ and $T$ 
duality)
of the various compactifications.

A related composite
picture of $p$-brane solutions is the bound states picture
\cite{bmike,bmirjam,bjoachim,bed,bsusy,bchris,barkady}.
For the simplest
case of extremal black hole solutions in $D=4$, consider the
Einstein-Maxwell-scalar action
\begin{equation}
S={1\over G_N} \int d^4x \left( R -
{1\over 2}(\partial\phi)^2 - {1\over
4}
e^{-a\phi} F_2^2 \right),
\end{equation}
where $a$ is an arbitrary parameter. It turns out that for the
specific values of $a=\sqrt{3},1,1/\sqrt{3}$ and $0$, 
supersymmetric
extreme black holes arising from string compactifications 
were found.
Moreover, from both the spacetime solutions 
\cite{bmirjam,bjoachim}\ and
supersymmetries \cite{bsusy}\ points of view, these 
four solutions
can be interpreted as bound states of $1,2,3$ and $4$ distinct
$a=\sqrt{3}$
black holes, respectively,
the latter corresponding to maximally supersymmetric
($N=4$ supersymmetry in an $N=8$, $D=4$ theory)
Kaluza-Klein or $H$-monopole solutions with flat
metric on moduli space \cite{rust}.
For example, the $a=0$ Reissner-Nordstr\"om
black hole arises as a bound state of two $T$-dual
pairs of electric/magnetic $a=\sqrt{3}$ black holes, each pair
producing
an $a=1$ black hole. Again, this feature extend quite 
naturally to
arbitrary supersymmetric $p$-branes in any dimension, 
as well as to
non-extremal, non-supersymmetric $p$-branes \cite{imirjam}.

As we shall see, this compositeness feature is crucial to 
the recent success
of string theory in reproducing the Beckenstein-Hawking 
\cite{Beck,Hawk}\ formula
for the entropy of black holes \cite{exfive,exfour,nonex}.
First, however, let us reconsider our quantum gravity 
discussion from the first lecture. We found that, 
for an elementary particle to exhibit an event
horizon and (possibly) behave like a black hole, it was
necessary that its Compton wavelength $\lambda_C$ be less
than its Schwarzschild radius $R_S$. A problem, however, arises
in this comparison, because most black holes are thermal
objects, and hence cannot reasonably be identified with
pure quantum states such as elementary particles. In fact,
in accordance with the laws
 of {\em black hole thermodynamics} \cite{Beck,Hawk}\ 
black holes radiate with a (Hawking) temperature constant
over the event horizon and proportional to the surface gravity:
$T_H\sim \kappa$. Furthermore, black holes possess an entropy
$S =A/4G_N$, where $A$ is the area of the horizon (the area
law), and 
$\delta A \geq 0$ in black hole processes. Finally, the energy
law of black hole thermodynamics takes the form
$dM=(\kappa/8\pi) dA + \Omega_H dJ$, in analogy with 
$dE=TdS-PdV$, where in the former case $\Omega_H$ 
is the angular velocity of the horizon and $J$ the angular 
momentum of the black hole.

Since an elementary particle is a pure state, it has entropy
$S=0$. According to the above, then, we need to first consider
black holes with zero area. For the specific four-dimensional
model above, the solutions with $a=\sqrt{3}$ and $a=1$
satisfy this requirement. Furthermore, their quantum numbers
(mass and charge) match those of particular supersymmetric
states of heterotic string theory \cite{bmike}. 
In addition, the dynamics of the black holes agree with the
four-point amplitudes of the corresponding string states 
in the low-velocity limit \cite{dynam}. For the $a=\sqrt{3}$
black holes and their corresponding states, this scattering
is trivial, while for the $a=1$ black holes and their 
corresponding states, we obtain Rutherford scattering. 
Of course this quantum number matching
and dynamical agreement does not
mean we can go ahead and identify the black hole
solutions with elementary string states, but at least we know
that we may possibly be on the right track.

Let us now consider black holes with $A\neq 0$. These should
now correspond to an {\em ensemble} of string states. Now
the laws of black hole thermodynamics follow from classical
general relativity. However, the laws of thermodynamics in 
general
follow from a microcanonical counting of quantum states,
{\em i.e.},  from statistical mechanics. An important test of 
a theory of quantum gravity is then the following: can one 
obtain the black hole laws of thermodynamics from a 
counting of microscopic quantum states, {\em i.e.}, is there
a quantum/statistical mechanical basis for these classical
laws? We are interested in performing this test for string
theory, where we have already developed a correspondence
between $p$-brane solutions and  elementary and solitonic
states. On the one side, we can construct a
black hole solution,
compute its area and deduce the entropy from $S=A/4G_N$.
On the other side, we can set up the ensemble of
states corresponding to this solution, compute its degeneracy
and take the logarithm to obtain the entropy.

Let us focus on a particular solution of the four-dimensional
model above, namely the case $a=0$, corresponding to the
(extremal)
Reissner-Nordstr\"om black hole. This solution is the bound
state of four constituent
$a=\sqrt{3}$ black holes. Let us embed this
solution in type IIB theory compactified from ten to
four dimensions. Then the four charges corresponding
to the four constituent black holes are given by (appropriately
normalized to represent number operators): $Q_1$, electric
with respect to $B_{\mu\nu}$, $Q_5$, magnetic with respect to 
$B_{\mu\nu}$,
$N$, electric with respect to $g_{5\mu}$ and 
$N'$, magnetic with respect to $g_{4\mu}$, 
where $x_4$ and $x_5$
are two compactified directions. $Q_1$ represents
the charge of a D-string, $Q_5$ that of a 
D-fivebrane and $N$ and $N'$ electric and magnetic Kaluza-Klein
black holes. Let us also simplify the picture by setting $N'=1$.

The classical solution has a nonzero area given by 
$A=8\pi G_N \sqrt{Q_1 Q_5 N}$, hence a Beckenstein-Hawking
entropy $S_{BH}=2\pi \sqrt{Q_1 Q_5 N}$. The setup of the 
corresponding
string states is the following.
We are interested in the case of large charges, 
corresponding to black hole solutions. For large $N$,
the number operator $N$ represents
the momentum of massless open strings going between
$Q_1$ D-strings and $Q_5$ D-fivebranes. The total number of
bosonic modes is then given by $4Q_1 Q_5$, since there is a
degeneracy associated with the extra $(6789)$ part of the 
compactified space. By supersymmetry, the number of fermionic
modes is also $4Q_1Q_5$.
This system is then like a 
$1+1$-dimensional gas of massless left 
moving particles with $4Q_1Q_5$ bosonic and fermionic species of
particles carrying total energy $N/R$, where $R$ is the 
radius of the circle. The number of such modes is 
given by $d(N) \sim \exp{2\pi \sqrt{Q_1 Q_5 N}}$, so that 
$S=\ln d(N)=2\pi \sqrt{Q_1 Q_5 N}=S_{BH}$, in agreement
with the area law. This is very exciting, as it is the first
time we can derive the area law from a quantum-mechanical
theory 
(string theory).\footnote{This argument is only
valid in the large $N$ limit, as otherwise the twisted
boundary conditions around the circle 
have to be taken into account. I am grateful to Juan Maldacena
for pointing this out.}

This agreement seems to hold for a number of cases.
In fact, for a given (static)
$p$-brane arising as a bound state
or as intersections of basic constituent $p$-branes of charge
$Q_1$,$Q_2$,$\ldots$,$Q_n$, the area law for the entropy yields
$S=A/4G = 2\pi \sqrt{Q_1 Q_2\ldots Q_n}$,
which again agrees, for large $Q_i$, with
the microscopic entropy formula obtained by counting
string states. This was first seen for five-dimensional extremal
black holes in \cite{exfive}\ and subsequently for 
four-dimensional
extremal black holes in \cite{exfour}. 
Analogous results for near-extremal
black holes were obtained in \cite{nonex}, 
which seems to indicate that
this sort of factorization is not a property of supersymmetry 
alone,
although it is only for supersymmetric solutions that one can
invoke non-renormalization theorems to protect the counting
of states in going from the perturbative state-counting picture
to the nonperturbative black hole picture.

This last argument is crucial: typically in string theory
the Schwarzschild radius $R_S \sim l_P g^m$, while 
$l_s \sim  l_P g^{-n}$, where $g$ is the string coupling,
$l_P$ and $l_s$ are the Planck
and string scale, respectively, and $m$ and $n$ are positive
integers (for $D=4$ heterotic string theory, for example,
$m=n=1$). So for $g \ll 1$, $R_S \ll l_P \ll l_s$, in
which case one has a perturbative state 
(or ensemble) but no black hole to
speak of (since the horizon is smaller than the string scale), 
while for  $g \gg 1$, $R_S \gg l_P \gg l_s$, 
so that one has a horizon and a black hole, but not a
perturbative string state. So unless we have an argument
that protects the counting of states in adiabatically tuning
the coupling from weak to strong, we have no right to assume 
that
the counting is preserved. Here we rely on
nonrenormalization due to suprsymmetry.

The question arises as to whether a similar 
factorization property holds
for arbitrary non-extremal black holes. Such a formula was, 
in fact,
found from the $p$-brane picture in \cite{nonex}, 
where, however, it was
noted that the corresponding D-brane counting argument was 
unknown.
More recently, it was argued in \cite{fat}\
that the counting arguments relating 
perturbative states to black
holes break down for ``fat'' black holes, {\it i.e.} large black
holes
far from extremality (such as the Schwarzschild black hole). 
However,
this does not in itself rule out the 
possibility of an analogous,
if not identical, compositeness
feature in the general case which will once again recover the
Beckenstein-Hawking formula and confirm an important success
of string theory as a theory of quantum gravity.
Nevertheless, these recent computations 
(as well the more recent work on greybody radiation \cite{GB})
have still not 
addressed the basic physical question, which is: what is
the physical process which takes an ensemble of states
into a black hole, and what do the quantum states look like in
the black hole (strong coupling) limit? 
These and other questions
remain to be answered, as well as the formulation of a 
duality-manifest string theory (M-theory and its
other single-alphabet cousins, such as F-theory \cite{Vafa}, 
should be thought of
as new names for nonperturbative string theory). 
In this regard,
connections with the better-understood Seiberg-Witten
\cite{SW}\ duality seem most promising \cite{Sen}.

\end{document}